\documentclass[a4paper,11pt]{amsart}
\usepackage{graphicx}
\begin{document}
\hyphenation{gra-vi-ta-tio-nal re-la-ti-vi-ty Gaus-sian
re-fe-ren-ce re-la-ti-ve gra-vi-ta-tion Schwarz-schild
ac-cor-dingly gra-vi-ta-tio-nal-ly re-la-ti-vi-stic pro-du-cing
de-ri-va-ti-ve ge-ne-ral ex-pli-citly des-cri-bed ma-the-ma-ti-cal
de-si-gnan-do-si coe-ren-za pro-blem gra-vi-ta-ting geo-de-sic
per-ga-mon}
\title[On spherically symmetric structures in GR]
{{\bf On spherically symmetric structures in GR}}

\author[Angelo Loinger]{Angelo Loinger}
\address{A.L. -- Dipartimento di Fisica, Universit\`a di Milano, Via
Celoria, 16 - 20133 Milano (Italy)}
\author[Tiziana Marsico]{Tiziana Marsico}
\address{T.M. -- Liceo Classico ``G. Berchet'', Via della Commenda, 26 - 20122 Milano (Italy)}
\email{angelo.loinger@mi.infn.it} \email{martiz64@libero.it}

\vskip0.50cm

\begin{abstract}
We reconsider some subtle points concerning the relativistic
treatment of the gravitational fields generated by spherically
symmetric structures.
\end{abstract}

\maketitle


PACS 0.40.20
-- General relativity.

\vskip0.50cm \noindent \emph{\textbf{Summary}}. -- \textbf{1}. On
the Einsteinian fields generated by spherosymmetrical bodies. The
continuity adjustments at the spherical boundary. -- \textbf{2}.
On Birkhoff's theorem. -- \textbf{2bis}. Physical spaces and
\emph{Bildr\"aume}-- \textbf{3}. Various forms of solution to the
problem of the Einsteinian field generated by a mass point.
Regular fields outside extended spherosymmetrical distributions of
matter. -- \textbf{4}. Event horizons and physical reality. --
\textbf{5}. and \textbf{5bis}. Geodesic motions of test particles
and light-rays in the Einsteinian field of a collapsed spherically
symmetric body $B$ with the minimal radius $(9/8)2m$.
Gravitational actions of a \emph{repulsive} kind. --
\textbf{5ter}. The event horizons are incapable of swallowing
anything. -- Appendices: Some observational consequences of sects.
\textbf{5}, \textbf{5bis}, \textbf{5ter}.

\normalsize

\vskip1.20cm \noindent \textbf{1}. -- As it is known \cite{1}, the
solution to the problem of the Einsteinian gravitational field
\emph{outside} a spherosymmetrical mass distribution at rest
(extended or point-like) is given -- if $r, \vartheta, \varphi$
are spherical polar coordinates -- by the following spacetime
interval:

\begin{eqnarray} \label{eq:one}
\textrm{d}s^2=\left[1-\frac{2m}{f(r)}\right]c^{2}\textrm{d}t^2-\left[1-\frac{2m}{f(r)}\right]^{-1}
[\textrm{d}f(r)]^2-[f(r)]^2\textrm{d}\omega^2 \quad;
\nonumber\\
 (\textrm{d}\omega^2\equiv
\textrm{d}\vartheta^2+\sin^{2}\vartheta \textrm{d}\varphi^2)
\quad,& &
\end{eqnarray}

where: $m\equiv GM/c^{2}$, $G$ is the gravitational constant; $M$
is the mass of the material distribution; $c$ is the speed of
light \emph{in vacuo}; $f(r)$ is any regular function of $r$, that
gives the Newtonian potential $GM/r$ for large values of $r$.
Eq.(\ref{eq:one}) holds \emph{only} for $r>2m$: indeed,  for
$f(r)\leq 2m$ the $\textrm{d}s^2$ loses its \emph{physically
essential pseudo}-Riemannian character. On the other hand, it is
not reasonable (as Marcel Brillouin and Nathan Rosen explicitly
remarked) to invert, within $f(r)<2m$, the roles of the radial and
temporal coordinates, thus rendering time dependent a static
metrical tensor: a quite unphysical result. Reality cannot be
changed by decree.

\par If we choose $f(r)\equiv r$, we obtain the customary \emph{standard
form} of solution, which was discovered (independently) by Droste,
Hilbert, and Weyl. According to a \emph{locus communis}, this form
would be dictated, as it were, by any intrinsically geometric
approach. In reality, in any approach of this kind one starts
invariably with the choice $r^{2}\textrm{d}\omega^2$ for the
angular term of $\textrm{d}s^2$, and this implies consequently the
usual, standard expression for the other terms.

\par When one investigates the field of an extended spherically
symmetric distribution of matter, one is also confronted with the
problem of the continuity adjustment between the internal and the
external values of the metric tensor $g_{jk}(x)$, $(j,k=0,1,2,3)$,
$[(x)\equiv (x^{0},x^{1},x^{2},x^{3})]$,  -- and possibly of the
continuity adjustment between the internal and the external values
of the derivatives $\partial g_{jk}/ \partial x^{\alpha}$,
$(\alpha=1,2,3)$, \emph{in perfect analogy with the Newtonian
theory}. In this more satisfactory treatment, the external value
of $f(r)$ does \emph{not} generally coincide with the value of
$f(r)$ corresponding to the mass point endowed with the same mass
$M$ of the extended distribution. This fact was clarified by
Schwarzschild in his second fundamental memoir on GR \cite{2},
which solves rigorously the problem of the Einsteinian field
generated by a homogeneous sphere of an incompressible fluid. The
continuity adjustments of $g_{jk}(x)$ \textbf{\emph{and}} of
$\partial g_{jk}/ \partial x^{\alpha}$ at the spherical boundary
tell us that eq.(\ref{eq:one}) is externally valid \emph{not} for
the function $f(r)\equiv [r^{3}+(2m)^{3}]^{1/3}$ characterizing
the \emph{original} Schwarzschildian form of solution for the
gravitational field of a mass point \cite{3}, but for the function
$f(r)\equiv (r^{3}+\varrho)^{1/3}$, where $\varrho$ is a constant
different from $2m$.

\vskip0.80cm \noindent \textbf{2}. -- Birkhoff's theorem -- i.e.,
the assertion that the $\textrm{d}s^2$ \emph{outside} of any
extended spherosymmetrical distribution of matter does \emph{not}
depend on a possible material motion which keeps the spherical
symmetry (for instance, a rhythmical pulsation of the sphere), and
satisfies accordingly eq.(\ref{eq:one}) -- is usually demonstrated
for the standard form $[f(r)\equiv r]$. It is however intuitive
that the theorem is true for \emph{any} choice of the function
$f(r)$. A trivial formal proof runs as follows. To be determinate,
let us consider the treatment given by Landau and Lifchitz in
sect.\textbf{97} of their book \cite{4}. As it is well known, it
is always possible to start from the following expression of the
$\textrm{d}s^2$ -- cf. eq.(97,2) of \cite{4}:

\begin{equation} \label{eq:two}
\textrm{d}s^2 = \exp \,[\nu(r,t)] \,c^{2}\textrm{d}t^2 - \exp \,
[\lambda(r,t)]\, \textrm{d}r^2 - r^2\textrm{d}\omega^2 \quad,
\end{equation}

which holds \emph{within and without} the material medium. The
functions $\lambda(r,t)$ and $\nu(r,t)$ will be determined by
Einstein equations. (Of course, with the above choice for the
angular part, Landau and Lifchitz will obtain the standard form of
$\textrm{d}s^2$ for the external region).

\par Now, put in eq.(\ref{eq:two}), in lieu of the usual polar
coordinate $r$, a generic function $f(r)$ of it, and call $u$ this
new radial coordinate. We remark immediately that all the
computations of sect.\textbf{97} of \cite{4} remain valid also in
this case.  Thus, one arrives at these results: \emph{i}) the
function $\lambda$ does not depend on time, and \emph{ii}):

\begin{equation} \label{eq:three}
\lambda(u) + \nu(u,t) = \textrm{a function } F(t) \quad ;
\end{equation}

it is easy to infer from eq.(\ref{eq:three}), by means of a
suitable change of time variable: $t\rightarrow t' =\psi(t)$,
that the \emph{external} field is always \emph{time independent}
and satisfies eq.(\ref{eq:one}). \emph{Q.e.d.} --

\vskip0.80cm \noindent \textbf{2bis}. -- One remarks usually that
the above choice (see eq.(\ref{eq:two})) $r^{2}\textrm{d}\omega^2$
for the angular term implies that the surface
$r=\textrm{constant}$ has the area
$A=4\pi(\textrm{constant})^{2}$: a \emph{Euclidean} formula! The
explanation is simple: this formula does \emph{not} give the
``natural'' expression of the area of the surface
$r=\textrm{constant}$, but its expression as measured in a
suitable three-dimensional \emph{Bildraum}, an auxiliary (and
physically fictitious) flat space. The difference between a
physical space and a ``picture space'' is conceptually essential.
However, few authors (e.g., Weyl \cite{2} and Fock \cite{5}) point
out explicitly this diversity.

\par Let us observe that in many instances the intervention of a
\emph{Bildraum} cannot be avoided -- and for a plain reason. For
clarity's sake, let us consider again Schwarzschild's problem. We
do not know \emph{a priori} the precise structure of the spacetime
manifold generated by our gravitating mass. Consequently, we are
not able to introduce a curvilinear coordinate system that is
``really adapted'' to the manifold geometry. In practice, we are
obliged to start with a coordinate system suggested by
simplicity's considerations, \emph{in primis} by the symmetry
properties of the problem.

 \vskip0.80cm \noindent \textbf{3}. -- The standard form of
 solution $[f(r)\equiv r]$, when considered for the field of a
 mass point, has a ``hard'' singularity at $r=0$ (i.e., a
 singularity for which Kretschmann's scalar is infinite) and a
 ``soft'' singularity at $r=2m$ (Kretschmann's scalar  of a finite
 value). (Of course, this form is physically and mathematically
 valid \emph{only} for $r>2m$, contrary to a diffuse belief.)

 \par It is instructive to compare the above form with other forms
 of solution, in particular with Fock's form, for which
 $f(r)\equiv r+m$ \cite{5}, and with Schwarzschild's \cite{3} and
 Brillouin's
 \cite{3} forms for which $f(r)\equiv [r^{3}+(2m)^{3}]^{1/3}$ and
 $f(r)=r+2m$, respectively. For a moment, and only for clarity's
 sake, let us call $r'$ the radial coordinate of standard form,
 and with $r''$ and $r'''$, respectively, the radial coordinates
 of Fock's form and of Schwarzschild's and Brillouin's forms.
 Fock's $r''$ has its origin ($r''=0$) at $r'=m$; Schwarzschild's
 and Brillouin's $r'''$ has its origin ($r'''=0$) at $r'=2m$. Thus
 the ``hard'' singularity at $r'=0$  of standards
 $\textrm{d}s^2$, which belongs to the unphysical region $0\leq r'
 \leq 2m$ that \emph{impairs} the pseudo-Riemannian character of
 the interval, has been removed. Fock's $\textrm{d}s^2$  holds
 only for $r''>m$. Schwarzschild's and Brillouin's
 $\textrm{d}s^2$'s hold only for $r'''>0$: \emph{they are} \textbf{\emph{maximally
 extended}}. It is evident that the above forms, considered for
 $r'>2m$, $r''>m$, $r'''>0$, respectively, are
 \emph{diffeomorphic}, and therefore mathematically and physically
 \emph{equivalent}.

 \par The known forms of solution by Lema\^itre (1933), Synge
 (1950), Finkelstein (1958), Kruskal and Szekeres (1960) are
 superfluous exertions; moreover, they make an essential use of
 coordinate transformations the derivatives of which are
 \emph{singular} at $r'=2m$ ``in just the appropriate way for providing a
 transformed metric that is regular there'' (Antoci and Liebscher, 2001). We
 observe that the singularity $r'=2m$ (or $r''=m$, or $r'''=0$)
 corresponds to the existence of a gravitating mass point -- and
 therefore it does \emph{not} represent a defect of the
 coordinate chart.

 \par It is not difficult to see that eq.(\ref{eq:one}) admits of
 infinite functions $f(r)$ such that the corresponding $\textrm{d}s^{2}$
 is \emph{everywhere regular for} $r\geq 0$ \cite{6}; e.g.,
 $f(r)\equiv r+3m$ (i.e., $r'\equiv r^{IV}+3m\geq 3m$).

 \par These regular solutions can be interpreted as representing
 the \emph{external} values of the $g_{jk}$'s of various
 spherically symmetric distributions with different structures and
 different radii. We shall see that a particularly significant
 solution of this kind is that for which $f(r)\equiv r+(9/8)2m$ --
 i.e., $r'\equiv r^{V}+(9/8)2m \geq (9/8)2m$.

 \vskip0.50cm \noindent \textbf{4}. -- The so-called
 \emph{event-horizons} corresponding to the singular geometric
 \emph{loci} of the various forms of solution for a mass point -- in
 particular, $r=2m$ for the standard form, $r=m$ for Fock's form,
 $r=0$ for Schwarzschild's and Brillouin's forms -- have a very
 dubious physical meaning, if considered with unprejudiced mind
 (se sect.\textbf{5ter}).

 \par We remark that: \emph{i}) as it can be rigorously
 demonstrated \cite{7}, the spherosymmetrical gravitational
 collapse of a massive (or supermassive) celestial body with a
 time-dependent pressure ends in a small structure endowed with a
 \textbf{\emph{finite}} volume; \emph{ii}) the minimal radius of
 spherically symmetric bodies (of a given mass $M$) consisting of
 perfect fluids at rest (in particular, of incompressible fluids
 \cite{2}) is equal to $(9/8)(2m)$, see \cite{2}, \cite{8}.

 \par Last, but not least, we wish to emphasize that it is not
 legitimate to hypothetize the existence of an event horizon in a
 celestial body which is a member of a binary system -- an
 assumption that many authors do. Indeed, as it was pointed out by
 McVittie many years ago (in 1972), an existence theorem would be
 needed to show that Einstein equations contain solutions which
 represent a binary system of stars having as a member a
 gravitating mass point. A simple analogy with Newton
 gravitational theory is \emph{not} sufficient.

\vskip0.50cm \noindent \textbf{5}. -- The precise meaning of
mentioned (see sect.\textbf{4}) minimal radius $(9/8)/2m$ is the
following: it is ``\emph{der au\ss en geme\ss ene Radius}''
$\textrm{P}_{a,\textrm{min}}$ (see \cite{2}) by means of which the
spherical volume $V_{a,\textrm{min}}$ of the considered material
distribution is measured in a given \emph{Bildraum} (not in the
real spacetime manifold!) by $(4/3)\pi
\textrm{P}_{a,\textrm{min}}^{3}$.

\par In the \emph{standard} coordinate system we have $\textrm{P}_{a}\equiv
r_{a}$, where $r_{a}$ is the radial coordinate of the points of
the spherical body.

\par The final volume $V_{\textrm{fin}}$ of a collapsed massive
star with time-dependent pressure and mass density, $p(t)$ and
$\varrho(t)$, is \cite{7}:

\begin{equation} \label{eq:four}
V_{\textrm{fin}} = \frac{4}{3} \, \pi \, \left[\,
\varepsilon_{0}^{2} \, (1+\varepsilon_{0})^{-2}\,\right]
\,r_{b}^{3} \quad,
\end{equation}

where $\varepsilon_{0} \equiv p(0) / (c^{2} \varrho\,
(\vartheta))$, and $r_{b}$ is the radial coordinate -- in a
Friedmann's coordinate system -- of the spherical boundary at $t=
t_{0}= 0$. Both $V_{a,\textrm{min}}$ and $V_{\textrm{fin}}$ are
defined in flat \emph{Bildr\"{a}ume}, which are \emph{in
abstracto} the same three-dimensional ``picture space''.
Accordingly, we can put $[\textrm{P}_{a,\textrm{min}} =
r_{a,\textrm{min}}= (9/8)2m]$:

\begin{equation} \label{eq:five}
\frac{4}{3} \, \pi \, [(9/8)2m]^{3} = \frac{4}{3} \, \pi \,
\left[\, \varepsilon_{0}^{2} \, (1+\varepsilon_{0})^{-2}\,\right]
\,r_{b}^{3} \quad,
\end{equation}

from which:

\begin{equation} \label{eq:six}
\frac{9}{4} \,m = \, \left[ \,\varepsilon_{0}^{2/3} \,
(1+\varepsilon_{0})^{-2/3}\,\right] \,r_{b} \quad;
\end{equation}

for a given mass $M$, suitable choices of $\varepsilon_{0}$ and
$r_{b}$ allow us to satisfy this equation.

\par By exploiting now some beautiful computations by Hilbert
\cite{9} of the geodesic lines in the spacetime manifold of a
gravitating \emph{mass point}, we shall exhibit a significant
diagram (see Fig. 1) which shows the squared velocity
$(\textrm{d}r/ c \textrm{d}t)^2$ of a test particle in radial
motion through the \emph{external} region of a spherically
symmetric \emph{body} of the minimal radius $(9/8)/2m$.

\par Hilbert begins with a fundamental remark: it is always
possible to find a suitable coordinate system for which
\emph{both} the gravitating centre \emph{and} the test particle
are \emph{at rest} \cite{10}: a consequence of ``plasticity'' of
the reference frames of GR. (Accordingly, the notion of affine
geodesic completeness of a manifold is not of primary importance
\emph{in GR}).

\par Then, Hilbert finds the most general expression of geodesic
lines, and investigates in detail the motions on the orbits
$r=\textrm{constant}$, and the radial ones.

\par The circular motions are restricted by the following
relations (Hilbert's $\alpha$ is equal to our $2m$):

\begin{equation} \label{eq:seven}
r > \frac{3}{2} (2m) > \frac{9}{8} (2m) \quad,
\end{equation}

\begin{equation} \label{eq:eight}
\frac{v}{c} < \frac{1}{\sqrt3} \quad,
\end{equation}

where $v=c(m/r)^{1/2}$ is the ordinary velocity (Hilbert puts
$c=1$). They are a striking consequence of spatio-temporal
curvature, which acts \textbf{\emph{repulsively}} for \emph{small}
values of the radial coordinate $r$.  Inequality (\ref{eq:seven})
represents a \emph{reinforcement} of $r>2m$, the validity
condition of standard $\textrm{d}s^2$. The meaning of inequality
(\ref{eq:eight}) is the following: when the coordinate $r$ of the
circular orbit decreases, the particle velocity tends to the
maximal value $c/\sqrt3$ -- in contrast with Newton theory, for
which this velocity increases illimitably, because there is no
restriction as inequality (\ref{eq:seven}).

\par Hilbert remarks that Schwarzschild's result \cite{3}
$v<c/\sqrt2$ has been obtained as a consequence of inequality
$r>2m$, which is weaker than relation (\ref{eq:seven}). (Of
course, Hilbert has made the formal translation from
Schwarzschild's coordinate frame to the standard one.)

\par Finally, Hilbert points out that the general equation of
motion of the test particle \cite{11} admits as solutions infinite
curves that approach indefinitely by spiraling every allowed
circular orbit -- as it is  required by Poincar\`e's general
theory of orbits \cite{12}. For the circular motions of light we
have a coordinate radius $r=(3/2)(2m)$ and a velocity
$v=c/\sqrt3$. There are infinite Poincar\`e's curves that approach
indefinitely by spiraling this circular trajectory.

\par For the radial motions of a test particle the action of the
spacetime curvature is quite peculiar, as we shall see. The
differential equation of these motions is ($2m<r<\infty$)
\cite{13}:

\begin{equation} \label{eq:nine}
\frac{1}{c^{2}} \frac{\textrm{d}^{2}r}{\textrm{d}t^{2}}-
\frac{3}{2} \, \frac{2m}{r(r-2m)} \left(
\frac{\textrm{d}r}{c\textrm{d}t}\right)^{2} +
\frac{m(r-2m)}{r^{3}}=0 \quad,
\end{equation}

with the following integral ($A$ is an integration constant with
\emph{negative} values; we have $(2/3)\leq |A| \leq 1$):

\begin{equation} \label{eq:ten}
\left(\frac{\textrm{d}r}{c\,\textrm{d}t}\right)^{2} =
\left(\frac{r-2m}{r}\right)^{2} + A
\left(\frac{r-2m}{r}\right)^{3} \quad.
\end{equation}

According to eq.(\ref{eq:nine}), the acceleration is negative
(gravitation acts attractively), or positive (gravitation acts
\emph{repulsively}) when, respectively:

\begin{equation} \label{eq:eleven}
\left| \frac{\textrm{d}r}{c\,\textrm{d}t}\right| <
\frac{1}{\sqrt3} \, \frac{r-2m}{r} \quad,
\end{equation}

or

\begin{equation} \label{eq:twelve}
\left| \frac{\textrm{d}r}{c\,\textrm{d}t}\right| >
\frac{1}{\sqrt3} \, \frac{r-2m}{r} \quad.
\end{equation}

It is instructive to consider the case $A=-1$; the test particle
starts from $r=\infty$ with zero velocity:
$\textrm{d}r/\textrm{d}t=0$.

\par If we set for brevity $x:= r/(2m)$ and
$y:=\left(\textrm{d}r / c\,\textrm{d}t\right)^{2}$, we see that
the function

\begin{equation} \label{eq:thirteen}
y(x)= \left(\frac{x-1}{x} \right)^{2} \left[ 1-
\frac{x-1}{x}\right] \quad; \quad 1<x<\infty \quad,
\end{equation}

reaches its maximum value $2^{2}/3^{3}$ at $x=3$:
$y(3)=2^{2}/3^{3}$. At $x=9/8$, we have $y(9/8)=2^{3}/3^{6}$;
$(9/8=1,125)$; $y(9/8)/y(3)=2/3^{3}$; $\sqrt{y(9/8)} =
\left(\textrm{d}r / c\,\textrm{d}t\right)_{r=(9/8)2m} = 2\sqrt2
/3^{3}\simeq 0,104757$. In Fig. 1 we give a diagram of function
(\ref{eq:thirteen}) for some significant values. We see the
impressive fall of particle velocity for $x<3$: a quite
``anti-Newtonian'' result!

\par For a radial motion of light, we have immediately from
$\textrm{d}s^2=0$ that

\begin{equation} \label{eq:fourteen}
\left| \frac{\textrm{d}r}{c\,\textrm{d}t}\right| = \frac{r-2m}{r}
\quad;  \quad (A=0) \quad;
\end{equation}

by virtue of inequality (\ref{eq:twelve}), eq.(\ref{eq:fourteen})
tells us that light is \textbf{\emph{repulsed}} by our extended
body. If a light-ray starts with velocity $c$ at $r=\infty$ , it
arrives at $r=(9/8)2m$ with velocity $(1/9)c$.

\par Finally, as Schwarzschild demonstrated \cite{2}, the light
arrives at the centre of the body of radius $(9/8)2m$ with a
velocity equal to \textbf{\emph{zero}} \cite{14}: an important
result for explaining the observational data concerning some X-ray
\emph{novae} \cite{15}.

\par Newtonian limit of eq.(\ref{eq:nine}): if $2m$ and the
particle velocity $\textrm{d}r/\textrm{d}t$ are small,
eq.(\ref{eq:nine}) is approximately equal to

\begin{equation} \label{eq:fifteen}
\frac{\textrm{d}^{2}r}{\textrm{d}t^{2}} = -\frac{c^{2}m}{r^{2}} =
- \frac{GM}{r^{2}} \quad.
\end{equation}

\vskip0.80cm \noindent \textbf{5bis}. -- If $A:= - |A|$,
$x:=r/(2m)$, $y:=(\textrm{d}r/c\,\textrm{d}t)^{2}$,
eq.(\ref{eq:ten}) can be written as follows:

\begin{equation} \label{eq:tenprime} \tag{10'}
y(x) = \left(\frac{x-1}{x}\right)^{2} \left[1-\,
|A|\frac{x-1}{x}\right] \quad, \quad (1<x<\infty) \quad.
\end{equation}

The value $x=x_{M}$ for which $y(x_{M})$ gives the maximal value
of $y(x)$ is:

\begin{equation} \label{eq:sixteen}
x_{M}= \frac{3|A|}{3|A|-2} \quad; \quad\left(\frac{2}{3}\leq |A|
\leq 1\right) \quad;
\end{equation}

if $|A|=2/3$, we have $x_{M}=\infty$, $y(x_{M})=1/3$, i.e.
$\textrm{d}r/\textrm{d}t=c/\sqrt3$. For \emph{any} value of the
mass of gravitating centre (extended or point-like), a test
particle which starts from $r=r_{M}=\infty$ with velocity
$v=c/\sqrt3$ must travel against a \emph{repulsive} gravitational
action. It will arrive at $r=(9/8)2m$ with the velocity
$v=(5/27)(c/\sqrt3)$

\par For $|A|=1$, we know (see sect.\textbf{5}) that the velocity
at $r=(9/8)2m$ is $v=(2\sqrt2/27)c$, if it had started from
infinite with a zero velocity; in this case, the gravitational
actions of Sun, planets, white dwarfs, and neutron stars are only
attractive.

\par For $|A|=0.8$, $x_{M}=6.0$; for $|A|=0.83$, $x_{M}=5.75$; for $|A|=0.89$,
$x_{M}=4.0$.

\par For $x_{M}=6.0$, $r_{M}=6\cdot 2\,m$ ; for a neutron star $2m \approx
5.3$ km, $r_{M}\approx 31.8$ km; the star radius $\textrm{P}_{a}
\approx 10$ km. Thus along $21.8$ km gravitation acts repulsively.

\vskip0.50cm \noindent \textbf{5ter}. -- The belief in the
physical significance of the event horizons encounters a great
difficulty. As it follows from previous sects.\textbf{5}.,
\textbf{5bis}, when the gravitating body is a \emph{mass point},
the test particles and the light rays do \emph{not} reach
generally in their motion the surface $r=2m$; only in the radial
motions (see eqs.(\ref{eq:nine}), (\ref{eq:ten}),
(\ref{eq:fourteen}), in particular) they can approach this
surface, but they arrive at it with velocities
$\textrm{d}r/\textrm{d}t$ and accelerations
$\textrm{d}^{2}r/\textrm{d}t^{2}$ which are equal to \emph{zero}.
(Of course, we have an identical conclusion if we describe the
Einsteinian field of the gravitating mass point with the forms of
solution, e.g., by Schwarzschild and Brillouin \cite{3} or by Fock
\cite{5}: test particles and light rays arrive at $r=0$, resp. at
$r=m$, with velocities and accelerations equal to zero.)

\par Accordingly, the event horizons are incapable of swallowing
anything! --

\par Unfortunately, in the current literature the classic memoirs
by Schwarzschild, Hilbert and Levi-Civita are ignored or
disfigured \cite{16}.

\newpage
\small \vskip0.5cm\par\hfill {\emph{``Soll man euch immer und
immer beplappern?}
  \par\hfill \emph{Gewinnt ihr nie einen freien Blick?''}
  \par\hfill \emph{Sie frieren, da\ss {} ihnen die Z\"ahne klappern}
       \par\hfill \emph{Das hei\ss en sie nachher Kritik.}
       \vskip0.10cm\par\hfill J.W. v. Goethe}

\normalsize \vskip0.80cm

\begin{center}
\noindent \small \emph{\textbf{APPENDIX A}}
\end{center}

\normalsize \noindent \vskip0.20cm
\par The title of the first paper quoted in \cite{15} is
``X-ray novae and the evidence from black hole event horizons''.
--  The abstract runs as follows: ``We discuss new observations of
X-ray novae which provide strong evidence that black holes have
event horizons. Optical observations of 13 X-ray novae indicate
that these binary stars contain collapsed objects too heavy to be
stable neutron stars. The objects have been identified as black
hole candidates. X-ray observations of several of these X-ray
novae in quiescence with the \emph{Chandra} X-ray Observatory show
that the systems are approximately 100 times fainter than nearly
identical X-ray novae containing neutron stars.  The
advection-dominated accretion flow $[$ADAF$]$ model provides a
natural explanation for the difference.  In this model, the
accreting gas reaches the accretor at the center with a large
amount of thermal energy. If the accretor is a black hole, the
thermal energy will disappear through the event horizon, and the
object will be very dim. If the accretor is a  neutron star or any
other object with a surface, the energy will be radiated from the
surface, and the object will be bright. $[\ldots]$.'' --

\par By virtue of the results of sects.\textbf{5}, \textbf{5bis},
\textbf{5ter}, the above explanation of the observational data by
means of the notion of event horizon is \emph{not} reasonable.

\par A simple explanation is obtained if we consider, in lieu of a
gravitating mass point, a gravitating small body $B$ of radius
$(9/8)2m$. As we have seen in sect.\textbf{5}, when a light-ray in
radial motion reaches $B$, it goes through $B$ and arrives at its
center with a velocity equal to zero. The accretor is now the body
$B$. The accreting gas in radial motion reaches $B$ with a large
amount of thermal energy, and with a remarkable speed. This very
hot material heats up star $B$. A considerable amount of the
energy will go through our accretor  as e.m. radiation, and will
arrive at its centre with a velocity equal to zero. In other
terms, a significant part of the radiation will not escape from
the stellar surface, but will be ``absorbed'' by object $B$, which
will be dim.

\par We remark that, since $(9/8)2m < (3/2)2m$, no Poincar\'e's
spiraling orbit of particle, or light-ray, can reach the surface
of $B$.

\vskip0.80cm
\begin{center} \noindent \small \emph{\textbf{APPENDIX
B}}
\end{center}

\noindent \vskip0.20cm
\par The title of the second paper quoted in
\cite{15} is ``X-ray QPOs in Black-Hole Binary Systems'', where
QPOs means ``quasiperiodic oscillations''. In the second paragraph
of the ``Introduction'' we read: ``Observations with the
\emph{Rossi} X-ray Timing Explorer (RXTE) have pioneered efforts
to further study black holes and their occasional relativistic
jets via broad-band X-ray observations during active states of
accretion. The X-ray timing and spectral properties convey
information about physical processes that occur near the black
hole event horizon, and one of the primary research goals is to
obtain constraints on the black hole mass and spire using
predictions of general relativity (GR) in the strong field
regime.''

\par This paper and the second review article quoted in \cite{15}
concern the event horizons of BH's \emph{with spin}, i.e. the
event horizons of Kerr's corpuscles \cite{17}. We shall give in a
next Note a reasonable explanation of the observational data
reported in these articles. \vskip0.80cm

\begin{figure}[!hbp]
\begin{center}
\includegraphics[width=1.0\textwidth]{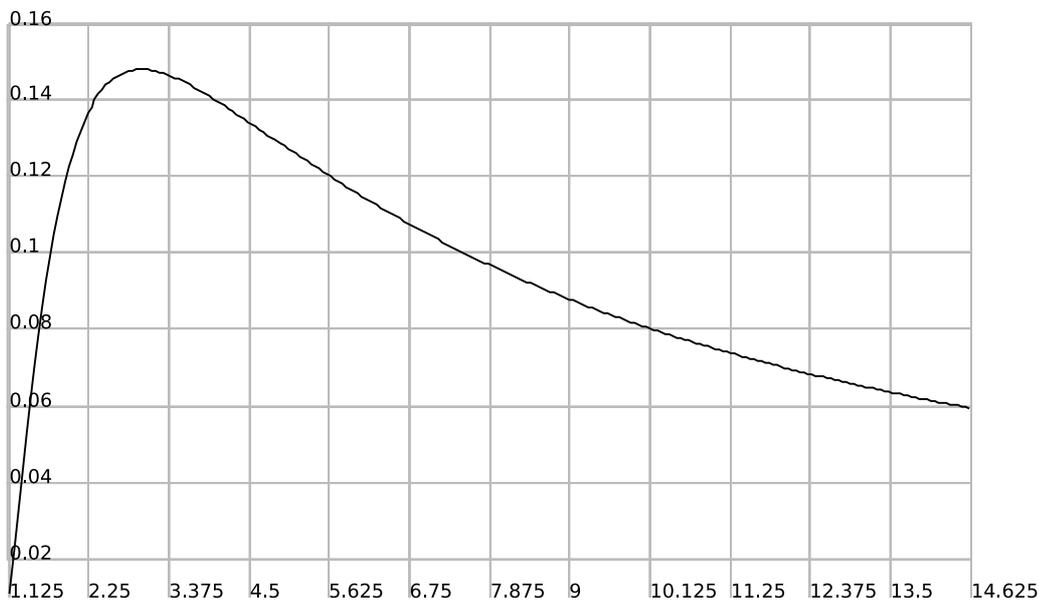}
\caption{Diagram of $y(x)=[(x-1)/x]^{2}[1-(x-1)/x]$ for some
values of $x$; $(9/8)\leq x <+\infty$; $\max(3.0,4/27)$;
$[y(9/8)]^{1/2}=2\sqrt2 /27$.}
\end{center}
\end{figure}


\begin{thebibliography}{99}
\bibitem{14}
For a trivial oversight, formula (44) of paper \cite{2} concerns
in reality $1/v$, and not $v$ (light speed).
\end{thebibliography}
\end{document}